\documentclass[]{article}
\usepackage{hyperref}
\usepackage{arxiv}

\usepackage{xspace}
\usepackage{amsmath}
\usepackage{amsfonts}
\usepackage{url}
\usepackage{graphicx}
\usepackage{xcolor}

\usepackage{subcaption}

\usepackage{tikz}
\usepackage{pgfplots}
\usetikzlibrary{calc,patterns,pgfplots.groupplots,positioning}

\newcommand{\NODE}{\ensuremath{n}\xspace}
\newcommand{\NODES}{\ensuremath{N}\xspace}

\newcommand{\ACK}{ACK\xspace}
\newcommand{\GACKIEEE}{GACK-IEEE\xspace}
\newcommand{\GACKBEACON}{GACK-Beacon\xspace}
\newcommand{\GACKCAP}{GACK-CAP\xspace}
\newcommand{\GACKGTS}{GACK-GTS\xspace}

\newcommand{\ACKWAIT}{macAckMaxWaitDuration\xspace}
\newcommand{\TURNAROUNDTIME}{\emph{aTurnaroundTime}\xspace}
\newcommand{\PAYLOAD}{\ensuremath{p}\xspace}

\newcommand{\BO}{\ensuremath{BO}\xspace}
\newcommand{\MO}{\ensuremath{MO}\xspace}
\newcommand{\SO}{\ensuremath{SO}\xspace}
\newcommand{\GAO}{\ensuremath{GAO}\xspace}

\newcommand{\SLOTDURATION}{\ensuremath{D_S}\xspace}
\newcommand{\SFDURATION}{\ensuremath{D_{SF}}\xspace}
\newcommand{\MSFDURATION}{\ensuremath{D_{MSF}}\xspace}
\newcommand{\BEACONDURATION}{\ensuremath{D_B}\xspace}
\newcommand{\SYMBOL}{\ensuremath{\mathcal{S}}\xspace}

\newcommand{\PGI}{\ensuremath{\tau}\xspace}

\usepackage{pgfplots}
\usepackage{xcolor}
\usetikzlibrary{patterns}

\colorlet{plotbordercolor}{black}
\definecolor{ploterrorcolor}{HTML}{AAAAAA}
\definecolor{plotcolor0}{RGB}{117, 189, 136}
\definecolor{plotcolor1}{RGB}{212, 114, 114}
\definecolor{plotcolor2}{RGB}{157, 167, 209}
\definecolor{plotcolor3}{RGB}{242, 179, 102}

\pgfplotscreateplotcyclelist{BarPlotCycleList}{
    {draw=plotbordercolor, fill=plotcolor0, postaction={pattern=none}},
    {draw=plotbordercolor, fill=plotcolor1, postaction={pattern=north east lines}},
    {draw=plotbordercolor, fill=plotcolor2, postaction={pattern=dots}},
    {draw=plotbordercolor, fill=plotcolor3, postaction={pattern=none}}
}

\pgfplotscreateplotcyclelist{LinePlotCycleList}{
    {color=plotcolor0, very thick, mark=none, solid},
    {color=plotcolor1, very thick, mark=none, densely dotted},
    {color=plotcolor2, very thick, mark=none, dashed},
    {color=plotcolor3, very thick, mark=none, dashdotted}
}

\pgfplotsset{
    compat=newest,
    width=\textwidth, 
    height=6cm,
    legend cell align = {left},
    xtick = data, 
    scaled x ticks = false,
    enlarge x limits = 0,
    xticklabel style = {
        /pgf/number format/fixed,
        /pgf/number format/precision=4,
    },
	ymin = 0,
	scaled y ticks = false,
    ylabel style = {
        align=center
    },
    yticklabel style = { 
        /pgf/number format/fixed, 
    }, 
    every axis plot post/.append style= {
        error bars/.cd, y dir=both, y explicit, error bar style={ploterrorcolor}
    },
    barplot/.style = {
        ybar, 
        bar width=.6cm,
        ymin=0,
        xtick=data,
        enlarge x limits=0.2,
        cycle list name = BarPlotCycleList,
    },
    lineplot/.style = {
        cycle list name = LinePlotCycleList,
    },
}

\title{Are Group Acknowledgements Worth Anything in IEEE 802.15.4 DSME: A Comparative Analysis}
\newcommand{\shorttitle}{Group Acknowledgements in IEEE 802.15.4 DSME: A Comparative Analysis}
\author{Florian Meyer, Phil Malessa, Jan Niklas Diercks, Volker Turau\\\\Institute of Telematics\\Hamburg University of Technology}

\begin{document}
\maketitle
	
\begin{abstract}
	For data collection scenarios in the Industrial Internet of Things, wireless communication provides a cost-effective and easy-to-deploy alternative to wired networks. The main focus lies on energy efficiency and reliability,  as many devices are battery operated. IEEE 802.15.4 DSME enhances reliability by acknowledging each packet individually, imposing an overhead for each transmitted packet, and increasing energy consumption. In networks with little interference, it may be beneficial to aggregate the acknowledgments for multiple nodes and broadcast them in a compressed format to all nodes in the neighborhood. The IEEE 802.15.4 2012 standard describes such a group acknowledgment scheme which, however, disappears in later iterations of the standard. This paper compares different group acknowledgment schemes and proposes a novel group acknowledgment scheme with the goal to examine whether group acknowledgments constitute a viable alternative to regular acknowledgments in reliable data-collection scenarios. Our analysis suggests that apart from a few cases, GACKs do not constitute a valid alternative to the direct acknowledgement of data packets. 
\end{abstract}

\section{Introduction} \label{sec:introduction}
Over the last few years, wireless communication experienced increased adoption in industrial applications, commonly known as \textit{Industrial Internet of Things} (IIoT), due to its ease of deployment and reduced setup cost in comparison to wired technologies \cite{Telematik_Kauer_2019_Diss}. The IEEE 802.15.4 standard and its extensions are specifically designed to cope with the requirements in this novel field of application, where the \textit{Deterministic and Synchronous Multichannel Extension} (DSME) increases throughput, reliability, and scalability by employing a TDMA/FDMA-based channel access and a distributed 3-way handshake for slot allocation \cite{ieee2012}. These slots, so-called \textit{Guaranteed Time Slots} (GTS), can be used for transmission of data packets.
Multiple packets can be transmitted per GTS \cite{meyer2020sending}, however, every packet is individually acknowledged imposing a significant overhead in terms of throughput and energy consumption. 


To increase the time a GTS can be used for sending application data, the acknowledgment overhead can be reduced, which allows for transmission of more data packets per GTS and reduces energy consumption. In the simplest form, data packets sent in a GTS are aggregated and acknowledged in a so-called \textit{group acknowledgement} (GACK) at the end of the GTS. This basic variant increases throughput in a trade-off for memory overhead, but can be further optimised by a more advanced variant in which data packets from multiple nodes are confirmed in a single broadcasted GACK. The IEEE 802.15.4 DSME 2012 standard defines such a GACK scheme, which was removed in later iterations of the standard \cite{ieee2012} - likely due to its complexity and inefficiency. We propose and investigate three novel possibilities to broadcast GACKs: in network beacons (\GACKBEACON), as management messages during a contention-based channel access phase (\GACKCAP), and in dedicated GTS (\GACKGTS).


The goal is to evaluate if GACKs constitute a viable alternative to regular acknowledgments in data collections scenarios with high reliability. In contrast to GACK schemes from the literature \cite{ieee2012, sahoo2017novel, battaglia2020novel}, the goal of the proposed GACK schemes is a higher flexibility. For this, retransmissions are conducted through regular GTS and the transmission of GACKs is not limited to coordinators, allowing every node in the network to utilize GACKs. 
Experiments are conducted using OMNeT++ \cite{varga2008overview} and indicate that GACKs can provide a 20\% higher reliability than regular ACKs in a best-case scenario. On the other hand, their performance is significantly worse than regular ACKs in all other scenarios so that we believe that GACKs do not constitute a valid alternative to direct acknowledgements. 

\section{Related Work}
Gomes et al. propose a hybrid channel hopping and channel adaptation scheme for DSME called H-DSME \cite{gomes2017comparison}. They identify the problem that GACKs are always transmitted on the same channel, which can potentially exhibit poor quality. Thus, H-DSME introduces a channel hopping mechanism for GACKs, where channels are selected using a round-robin mechanism. They compare H-DSME with two other techniques they propose, but unfortunately the results do not indicate the performance of GACKs in comparison to regular ACKs.  

DSME's reliability and end-to-end delay using GACKs according to the IEEE 802.15.4e 2012 standard are compared to IEEE 802.15.4 TSCH by Alderisi et al. in realistic and increasingly dense process automation environments \cite{alderisi2015simulative}. DSME yields a higher delay than TSCH for networks with less than 30 nodes but outperforms TSCH at 50 nodes, where GACKs had to be disabled to increase throughput and allow every node to allocate sufficiently many GTS. Alderisi et al. conclude that DSME's GACKs, where packets are retransmitted at the end of a superframe, severly increase the delay of regular data transmissions, so their results provide a good indication of the inflexibility of GACKs according to the IEEE 802.15.4 standard. 

Sahoo et al. propose a GACK scheme for IEEE 802.15.4e DSME where GACKs are transmitted as parts of regular beacons limiting their use to coordinators \cite{sahoo2017novel}. Additionally, they devise a retransmission scheme where lost packets are retransmitted in the CAP without using CSMA/CA, thus, reducing delay. The details of their work are further discussed in Sect.~\ref{sec:gack_beacon}.  

Similarly to \cite{sahoo2017novel}, Battaglia et al. propose a novel \ACK scheme based on the periodicity of traffic flows to prevent the retransmission overhead of GACKs according to the IEEE 802.15.4 standard \cite{battaglia2020novel}. In DSME, a GTS is repeated every MSF even if a flow has a period of $n$ MSFs, resulting in $n-1$ unused GTS. Their idea is to utilize these unused GTS to transmit $n-1$ replicas of a data message in the flow, giving the receiver $n-1$ additional opportunities to receive a message correctly. Obviously, the proposed scheme increases reliability though redundant information but greatly increases energy-consumption.  

\section{Overview of IEEE 802.15.4 DSME}
\begin{figure}[h]
	\centering
	\includegraphics{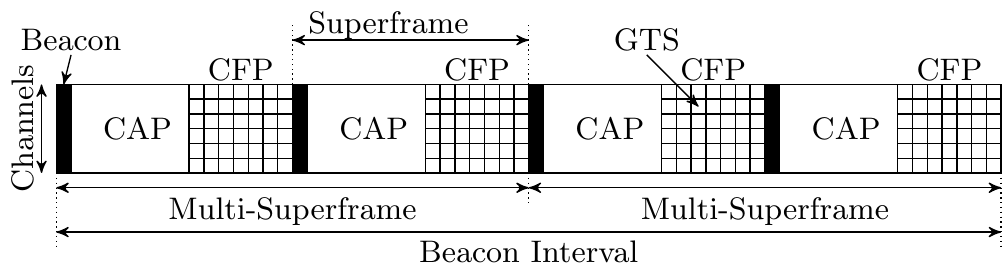}
	\caption{Frame structure of IEEE 802.15.4 DSME with $\MO-\SO = 1$ and $\BO-\MO=1$.}
	\label{fig:dsme_frame_structure}
\end{figure}

In IEEE 802.15.4 DSME \cite{ieee2020}, time is divided into \emph{superframes} (SF), which are further subdivided into 16 distinct time slots. The first of these is used for the transmission of beacons, conveying network and time information. The remaining 15 time slots are split into a \emph{contention-access period} (CAP) and a \emph{contention-free period} (CFP) consisting of 8 and 7 time slots, respectively. The CFP comprises a total of $7 * 16 = 112$ GTS spread over time and 16 frequencies, which allow, after initial allocation, contention-free communication between a pair of nodes. The (de)allocation of GTS is conducted in the CAP using a distributed 3-way handshake via CSMA/CA. Here, always the same channel is used. Multiple SFs are combined to a \emph{multi-superframe} (MSF), after which a schedule of allocated GTS repeats. DSME's complete frame structure is depicted in Fig.~\ref{fig:dsme_frame_structure}.

The 3-way GTS allocation handshake is initialized by the unicast transmission of a GTS-request from a node A to a node B. B responds by broadcasting a GTS-response to inform all nodes in its neighbourhood about the GTS to be allocated. When A receives the GTS-response, it finalizes the handshake with a broadcasted GTS-notify to also inform all nodes in its neighbourhood. If any of A's or B's neighbours have already allocated the GTS, i.e., a duplicate allocation occurs, or if any of the 3 messages is lost, the GTS allocation is rolled back using the same 3-way handshake.

DSME's frame structure is configured using three parameters: \emph{superframe order} (\SO), \emph{multisuperframe order} (\MO), and \emph{beacon order} (\BO). \SO determines the time slot duration $\SLOTDURATION = 0.96\ ms * 2^{\SO}$ and, thus, the superframe duration as $\SFDURATION = 16 * \SLOTDURATION$. The multisuperframe duration is given by \MO as $\MSFDURATION = 15.36\ ms * 2^{\MO} = \SFDURATION * 2^{\MO-\SO}$, where $2^{\MO-\SO}$ also determines the number of SFs per MSF. At last, the duration of a \textit{beacon interval} (BI) is given by $\BEACONDURATION = 15.36\ ms * 2^{\BO}$. The selection of these parameters is highly application-specific. For example, a larger BI allows for more network participants but also increases association time. 

To increase reliability, packets transmitted during a GTS are directly acknowledged. For this, nodes wait for a maximum of an \textit{\ACKWAIT}~$= 54$~symbols for an \emph{acknowledgment frame} (\ACK) and retransmit the data packet if the \ACK did not arrive in time. In a previous work we have shown that it is feasible and beneficial to transmit multiple packets per GTS \cite{meyer2020sending}.

\section{Group Acknowledgments}
In the first iteration of IEEE 802.15.4 DSME from 2012, GACKs were defined as part of the official standard \cite{ieee2012}. They have been removed in later iterations but their potential to increase DSME's throughput and energy-efficiency in reliable environments has yet to be fully determined. 
Several works propose new GACK schemes achieving auspicious results \cite{sahoo2017novel, battaglia2020novel}. The following sections provide an overview of existing GACK schemes (Sect.~\ref{sec:gack_ieee} and \ref{sec:gack_beacon}) and introduce three novel schemes \GACKBEACON, \GACKCAP and \GACKGTS (Sects.~\ref{sec:gack_beacon_our}, \ref{sec:gack_cap} and \ref{sec:gack_gts}) that will be compared throughout this work.

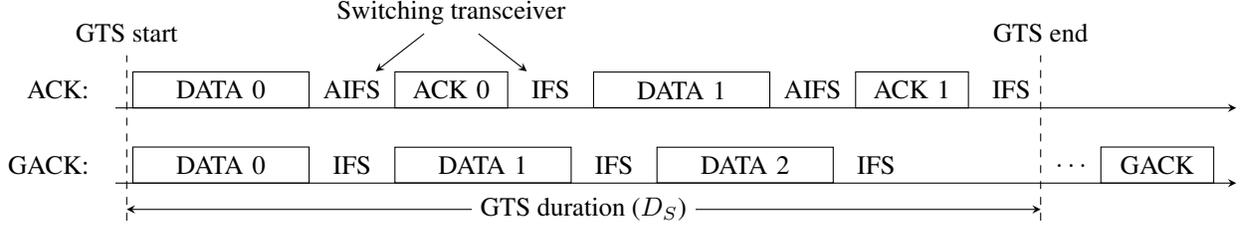
\begin{figure}
	\centering
\newcommand{\data}[4]{\node[right = of #2, draw, rectangle, text width=#4] (#1) {DATA #3}}
\newcommand{\aifs}[2]{\node[right = of #2, rectangle, text width=25] (#1) {AIFS}}
\newcommand{\ack}[3]{\node[right = of #2, draw, rectangle, text width=36] (#1) {ACK #3}}
\newcommand{\ifs}[2]{\node[right = of #2, rectangle, text width=25] (#1) {IFS}}
\newcommand{\cont}[2]{\node[right = of #2, rectangle] (#1) {$\dots$}}

\begin{tikzpicture}[node distance=0pt, align=center]
	\draw[-stealth] (-0.2, 0) -- (14.7, 0); 
	\node at (-0.1,0.24) (start0) {};
	\node[left=0.2cm of start0] {ACK:};
	\data{data0}{start0}{0}{60};
	\aifs{aifs0}{data0};
	\ack{ack0}{aifs0}{0};
	\ifs{ifs0}{ack0};
	\data{data1}{ifs0}{1}{60};
	\aifs{aifs1}{data1};
	\ack{ack1}{aifs1}{1};
	\ifs{ifs1}{ack1};
	
	\node[above=0.5cm of ack0] (switch) {Switching transceiver};
	\draw[-stealth] (switch) -- (aifs0);
	\draw[-stealth] (switch) -- (ifs0);
	
	
	\draw[-stealth] (-0.2, -1) -- (14.7, -1); 
	\node at (-0.1,-0.76) (start0) {};
	\node[left=0.2cm of start0] {GACK:};
	\data{data0}{start0}{0}{60};
	\ifs{ifs0}{data0};
	\data{data1}{ifs0}{1}{60};
	\ifs{ifs1}{data1};
	\data{data2}{ifs1}{2}{60};
	\ifs{ifs2}{data2};
        
    \node[right = 1.7cm of ifs2, rectangle] (gackdots) {$\dots$};     
    \node[right = of gackdots, rectangle, draw, minimum width=1.5cm] {GACK};
        
    \draw[dashed] (-0.05, -1.5) -- (-0.05, 0.7);
    \path[stealth-stealth] (-0.05,-1.35) edge node[fill=white] {GTS duration (\SLOTDURATION)} (12.1, -1.35);
    \draw[dashed] (12.1, -1.5) -- (12.1, 0.7);
    \node[align=center] at (-0.05, 1.0) {GTS start};
    \node[align=center] at (12.1, 1.0) {GTS end};
\end{tikzpicture}
	\caption{Transmission process using GACKs and regular ACKs within a single GTS with a MAC payload of 1 Byte.}
	\label{fig:gack_frame_time}
\end{figure}

In general, GACKs avoid the overhead of an acknowledgment and an \textit{acknowledgment inter frame space} (AIFS) for every data transmission, as illustrated in Fig.~\ref{fig:gack_frame_time}. This allows for more data transmissions within a single GTS and also avoids switching the transceiver from transmitting to receiving after every data transmission. 

\subsection{IEEE 802.15.4e 2012} \label{sec:gack_ieee}
\begin{figure}
	\centering
	\includegraphics[scale=1.1]{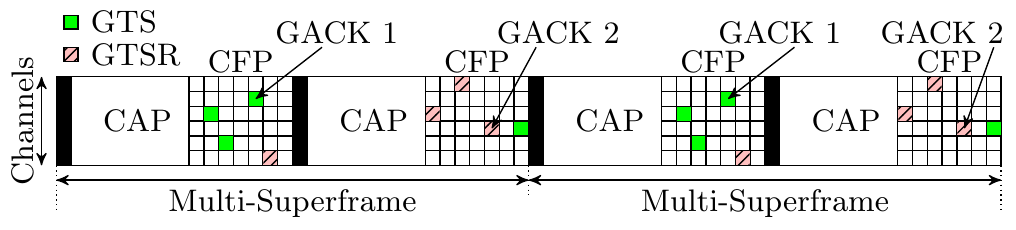}
	\caption{DSME's GACK scheme as defined by the 2012 amendment to the IEEE 802.15.4 standard.}
	\label{fig:gack_ieee}
\end{figure}

In the 2012 amendment to the IEEE 802.15.4 standard \cite{ieee2012}, GACKs were defined as part of DSME. Here, every coordinator allocates two GTS for the transmission of GACKs (GACK~1 and GACK~2) and announces their superframe ID, slot ID, and channel ID in its enhanced beacons. Thereby, all packets received before GACK~1 but after GACK~2 are acknowledged in GACK~1 while all packets received after GACK~1 but before GACK~2 are acknowledged in GACK~2. As shown in Fig.~\ref{fig:gack_ieee}, all GTS before GACK~1 are used for regular data transmission but for every allocated GTS another so-called \emph{GTS for retransmission} (GTSR) must be allocated between GACK~1 and GACK~2. These GTSR allow for retransmission of lost data packets. 

IEEE 802.15.4e specifies the frame format for a GACK, in which the reception status of a packet is indicated by a single bit in a bitmap. The size of the bitmap can be configured in bytes using two bits, for a maximum size of 4 bytes, i.e., 32 acknowledgeable packets. 

There are several shortcomings in the described GACK scheme: First, only coordinators can utilize GACKs because information about them is disseminated in their beacons. Additionally, many GTSR are unused if data packets are transmitted successfully, reducing potential throughput \cite{sahoo2017novel}. At last, the maximum size of the bitmap is too small for high-traffic scenarios. 

\subsection{Transmitting GACKs through beacons} \label{sec:gack_beacon}
\begin{figure}
    \centering
    \includegraphics[scale=1.1]{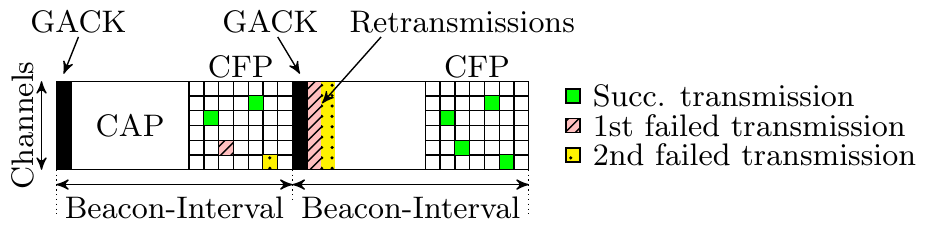}
    \caption{Operation of the GACK scheme according to Sahoo et al.} 
    \label{fig:gack_beacon}
\end{figure}

Due to the shortcomings of IEEE 802.15.4e's GACK scheme, Sahoo et al. propose a scheme where GACKs are transmitted as part of beacon messages \cite{sahoo2017novel}. For this, they extend every beacon with a 7 bit bitmap, where every bit indicates the successful transmission during a respective CFP slot, of which there are 7 CFP slots without CAP-reduction. On reception of a GACK, every node counts the number of lost packets $l$ which are then retransmitted in the first $l$ CAP slots without using CSMA/CA. During this time no other node is allowed to transmit a packet. Afterwards all nodes utilize the CAP normally. The operation of the scheme is depicted in Fig.~\ref{fig:gack_beacon}.

Even though the scheme by Sahoo et al. significantly increases throughput, it introduces a number of different drawbacks, e.g., the fixed bitmap size of 7 bits only allows for $\SO = \MO = \BO$, i.e., exactly 7 CFP slots per BI, and CAP-reduction is not usable. Similar to the IEEE 802.15.4 standard, GACKs are limited to coordinators only, disallowing upstream traffic. Additionally, retransmitting packets in the CAP can become a problem in scenarios with fluctuating traffic where many management messages are transmitted.

\subsubsection{\GACKBEACON} \label{sec:gack_beacon_our}
Although there are many shortcomings in the scheme described by Sahoo et al., the idea of transmitting GACKs in beacon frames seems to be promising as it does not restrain throughput in the CFP. Thus, we adopt this idea but combine it with a more flexible bitmap format, described in Sect.~\ref{sec:gack_bitmap}, and handle retransmissions through regular GTS transmissions instead of utilizing the CAP. This way, \SO, \MO, and \BO can be selected without any restrictions. In the following, this scheme is called \GACKBEACON. 

\subsection{\GACKCAP} \label{sec:gack_cap}
DSME's CAP is primarily used for transmitting management and broadcast messages so that its utilization in a settled network is usually low. Thus, the idea of \GACKCAP is to send GACKs as regular CAP messages. The main advantage is that GACKs impose no additional overhead during the CFP maximizing potential throughput. On the other hand, the scheme suffers from the same disadvantages as CSMA/CA, i.e., low reliability and high delays in dense networks with fluctuating traffic conditions \cite{meyer2019performance}. Similar to \GACKBEACON, retransmissions are performed in regular GTS. 

\subsection{\GACKGTS} \label{sec:gack_gts}
\begin{figure}
    \centering
    \includegraphics[]{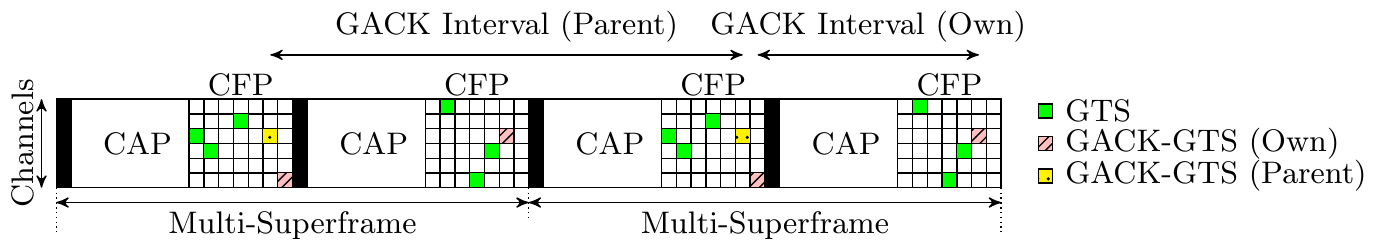}
    \caption{GACKs using the \GACKGTS scheme}
    \label{fig:gack_gts}
\end{figure}

The \GACKGTS scheme is not only a generalization but also a simplification of the \GACKIEEE scheme. The main idea is to allocate dedicated \GACKGTS during the CFP in which GACKs are broadcasted to nodes in the neighborhood. The frequency of the \GACKGTS, i.e., the GACK interval is determined by the \textit{group acknowledgement order} (\GAO), where $2^{\MO-\GAO}$ gives the number of \GACKGTS per MSF. An example for the whole scheme is shown in Fig.~\ref{fig:gack_gts}, where $\GAO=\MO$ is  used by a parent node and $\GAO-1=\MO$ locally. 

Allocation of the \GACKGTS is conducted with the regular 3-way handshake using an additional flag, i.e., when a node sends a GTS-request to allocate a normal GTS. The receiver of the GTS-request checks if there is already an \GACKGTS within one GACK interval of the GTS to be allocated and either selects the next \GACKGTS or allocates a new \GACKGTS within one GACK interval. The \GACKGTS is communicated with the requesting node in the GTS-response. Thereby, \GACKGTS are randomly allocated from the tail of a superframe while regular GTS are allocated randomly from the beginning. 

One should notice that in the described GACK allocation procedure, a node might return a \GACKGTS that conflicts with an already allocated GTS of the requesting node. There is no way to avoid such collisions as the \GACKGTS is communicated to several nodes and a change of the \GACKGTS would likely result in conflicts at other nodes. Thus, in case of a conflict, the requesting node can simply abort the handshake and try to allocate the GTS in another superframe to retrieve another \GACKGTS. If this fails, it can simply relocate the conflicting GTS to make room for the \GACKGTS, e.g., using a GTS-change command. 

\subsection{GACK bitmap format} \label{sec:gack_bitmap}
As described in Sect.~\ref{sec:gack_ieee}, IEEE 802.15.4e 2012 defines a bitmap structure for acknowledgments through GACKs, which supports up to 32 ACKs. Especially in scenarios with a large \SO where many packets can be transmitted per GTS \cite{meyer2020sending}, a larger and more flexible bitmap format is required. 

The proposed bitmap format is shown in Tbl.~\ref{tbl:gack_bitmap}. Unlike the GACK formats from Sect.~\ref{sec:gack_ieee} and Sect.~\ref{sec:gack_cap}, the bits in the bitmap do not directly address a GTS but acknowledge data packets relative to a specified sequence number. This prevents two issues: first, the bitmap does not contain one bit for each packet that might be sent, so it is smaller on average. Second, and more importantly, consistency problems are avoided because each bit is associated with a given sequence number. It would otherwise be impossible to determine which packet within a GTS failed. The GACK contains a \textit{payload} for each node from which a packet was received. This payload consists of the \textit{node address} identifying the node, the \textit{bitmap length}, the \textit{sequence number} of the first packet to be acknowledged, and a \textit{bitmap} indicating all packets received relative to that sequence number.

\begin{table}
    \centering
    \caption{Generalized GACK bitmap structure using \GACKBEACON, \GACKCAP and \GACKGTS.} 
    \label{tbl:gack_bitmap}
	\begin{tabular}{|c|c|c|c|c|c|c|c|c|}
        \hline
		\textbf{octets} & 1 & 2 & 1 & 1 & \textit{bitmap length} & ... & ... & ... \\\hline
		\textbf{field} & \# payloads & node addr. & bitmap length & seq. number & bitmap & ... & ... & ... \\\hline
		\textbf{group} & header & \multicolumn{4}{c|}{payload 1} & payload 2 & ... & payload n \\\hline
	\end{tabular}
\end{table}

Note that the proposed bitmap format is not yet optimal and its size can be further compressed with additional effort, e.g., by omitting the \textit{\# payloads} field \cite{chen2018gas}. However, the format is flexible and can be easily processed by sensor nodes with limited resources, making it well suited for the evaluated scenarios.

\section{Comparison and Hypotheses} 
To evaluate the performance of GACKs in IEEE~802.15.4~DSME, serveral metrics can be utilized. These include but are not limited to throughput, energy consumption, dwell time, \ACK delay and memory overhead. Due to the diversity of the presented \ACK schemes, it is impossible to derive general statements about their performance. Therefore, Tbl.~\ref{tbl:ack_performance} gives an indication of their expected performance for mentioned metrics. 

\begin{table}
	\centering
	\caption{Performance of \ACK schemes in IEEE 802.15.4 DSME.}
	\label{tbl:ack_performance}
	\begin{tabular}{r|c|c|c|c}
		& \ACK & \GACKBEACON & \GACKCAP & \GACKGTS \\\hline
		throughput 			& - & + & + & o \\
		energy consumption 	& - & o & + & o \\
		dwell time 			& o & + & + & o \\
		\ACK delay 			& + & - & - & o \\
		memory overhead 	& o & - & - & - \\
	\end{tabular}
\end{table}

In general, all GACK schemes are expected to provide higher throughput than the normal \ACK scheme, however, while \GACKCAP and \GACKBEACON provide the maximum throughput, \GACKGTS's throughput is slightly lower because it requires at least one GTS per coordinator for the transmission of GACKs. The dwell time directly depends on the throughput. On the other hand, \GACKBEACON and \GACKGTS exhibit a variable \ACK delay by tuning \BO and \GAO, respectively. \GACKCAP provides a relatively low expected \ACK delay of half the number of GTS per superframe, which also constitutes the lower bound for \GACKBEACON. \GACKGTS can reach a lower \ACK delay for a direct trade-off with throughput. It should be noted that the ACK delay is of secondary importance in reliable environments and is only relevant in case of packet loss. For all GACK schemes, the energy consumption is inversely proportional and memory overhead is proportional to the \ACK delay since more data packets are received and have to be stored in the GACK-bitmap before a GACK is transmitted. In particular, \GACKCAP requires the fewest memory while more is required for \GACKCAP and \GACKBEACON in its normal configuration. All GACK schemes are expected to require less energy and more memory than a normal \ACK. 


\section{Theoretical Considerations} \label{sec:theoretical}
Depending on the scenario, the goal of the MAC layer is to increase the possible throughput for the application - either in terms of transmittable packets or in terms of goodput. Thus, the following two sections, Sect.~\ref{sec:theoretical_packets} and \ref{sec:theoretical_goodput}, provide theoretical estimations of the maximum packet throughput and maximum goodput for different \ACK schemes. Together, they allow to assess the performance benefit of GACKs over regular ACKs. 

\subsection{Maximum Throughput} \label{sec:theoretical_packets}
As already mentioned in Sect.~\ref{sec:introduction}, GACKs increase the maximum number of transmittable packets per GTS. That is due to two reasons: First, there is no need to transmit an \ACK in response to every data packet. Furthermore, the IEEE 802.15.4 standard requires an \emph{interframe space} (IFS) for processing after every received data packet, which includes an \TURNAROUNDTIME for switching the transceiver from the RX to the TX state and vice versa. However, this is unnecessary if no \ACK is transmitted. 

The expected maximum throughput in packets per second $\mathbb{T}_{A|G}$ for a given \SO and payload \PAYLOAD in bytes is given by 
\begin{align}
	\mathbb{T}_{A|G} = \frac{\SYMBOL_{sec}}{\SYMBOL_{SF}} * GTS_{SF} * \lfloor \frac{\SYMBOL_{GTS}}{\SYMBOL_{A|G}(\PAYLOAD)} \rfloor, \label{eq:theoretical_packets_max_packets}
\end{align}
where $\SYMBOL_{sec} = 62500\ symbols$ is the constant symbol rate, $GTS_{SF}=7$ is the number of GTS per superframe, $\SYMBOL_{GTS} = 60 * 2^{\SO}\ symbols$ is the number of symbols per GTS and $\SYMBOL_{SF} = 16 * \SYMBOL_{GTS}$ is the number of symbols per SF. Finally, $\SYMBOL_{A}(\PAYLOAD)$ and $\SYMBOL_G(\PAYLOAD)$ are functions calculating the number of symbols per data transmission dependent on \PAYLOAD with and without \ACK and are given by 
\begin{align} 
	\SYMBOL_{A}(\PAYLOAD) &= \SYMBOL_H + 2\PAYLOAD + \SYMBOL_{\ACK} + \SYMBOL_{IFS}, \\ 
	\SYMBOL_{G}(\PAYLOAD) &= \SYMBOL_H + 2\PAYLOAD + \SYMBOL_{IFS} - \TURNAROUNDTIME, 
\end{align}
respectively, Here, \PAYLOAD is the payload length in bytes, $\SYMBOL_{\ACK}=34$ symbols is the overhead for the \ACK packet and an \emph{acknowledgment IFS} (AIFS), and $\SYMBOL_{IFS}$ is either a \emph{short inter frame space} (SIFS) with 12 symbols or a \emph{long inter frame space} (LIFS) with 40 symbols if $\PAYLOAD > 18$ bytes. At last, $\SYMBOL_H=34$ symbols is the header overhead for the data packet.  

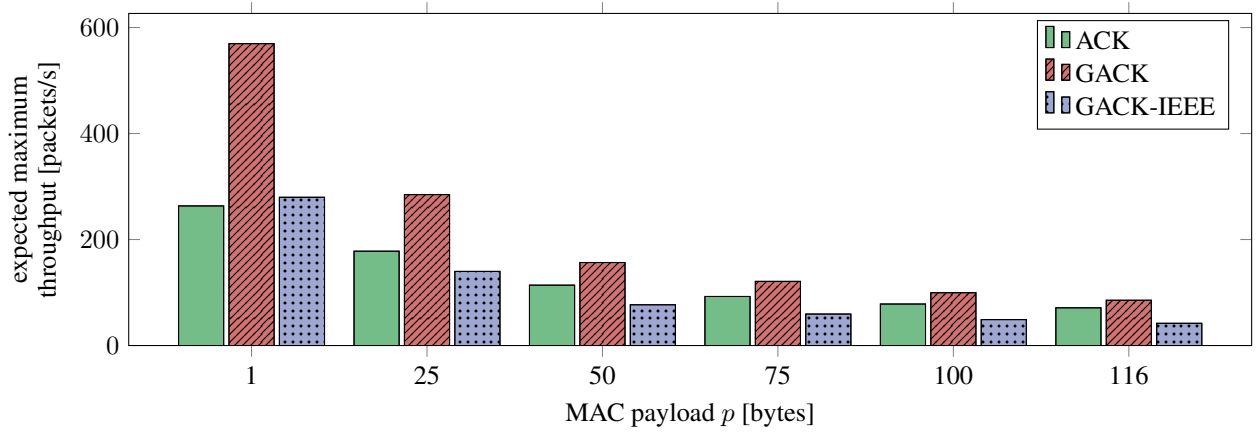
\begin{figure}
	\centering
	\begin{tikzpicture}
\begin{axis}[
    barplot, 
    xlabel = {MAC payload \PAYLOAD [bytes]},
    ylabel = {expected maximum\\throughput [packets/s]},  
    legend entries = {ACK, GACK, \GACKIEEE, \GACKGTS},
    legend columns = 1,
    symbolic x coords = {1, 25, 50, 75, 100, 116},
    enlarge x limits = 0.14,
]

\foreach \x in {ack, gack, gack_ieee} {
    \addplot+ table[x=payload, y=packets_\x, col sep=comma] {plots/theoretical/packets_per_second.csv};
}

\end{axis}
\end{tikzpicture} 
	\caption{Expected maximum packet rate using GACKs and regular ACKs for different payload length \PAYLOAD.}
	\label{fig:theoretical_packets}
\end{figure}

Fig.~\ref{fig:theoretical_packets} shows the expected maximum throughput of GACKs ($\mathbb{T}_G$) in comparison to regular ACKs ($\mathbb{T}_A$) for different values of \PAYLOAD for $\MO=8$. As one can see, the number of transmittable packets can be significantly increased for small \PAYLOAD using the proposed GACK schemes while for large \PAYLOAD no significant increase is achievable. Over all \SO, the expected maximum throughput for a given \PAYLOAD is almost constant as a superframe with a shorter length is repeated multiple times per second. As expected, the expected maximum throughput of \GACKIEEE is significant lower than for the other techniques because twice as many GTS have to be allocated for every transmission. \GACKIEEE only outperforms regular for $\PAYLOAD=1 Byte$. 

\subsection{Maximum Goodput} \label{sec:theoretical_goodput}
Contrasting Sect.~\ref{sec:theoretical_packets}, an application is not only interested in the raw packet throughput but rather in the goodput, i.e., the total application layer payload the MAC protocol can transmit per second. Thereby, the goodput is adjusted for header, acknowledgement and management overhead. 

The expected maximum goodput per second $\mathbb{G}_{A|G}$ in bytes is calculated similarly to Eq.~\ref{eq:theoretical_packets_max_packets} in Sect.~\ref{sec:theoretical_packets} as 
\begin{align}
	\mathbb{G}_{A|G} = \frac{\SYMBOL_{sec}}{\SYMBOL_{SF}} * GTS_{SF} * G_{A|G},
\end{align}
where $G_A$ and $G_G$ are the goodputs per GTS with and without ACKs, respectively. They are calculated as  
\begin{align}
	G_A &= 116 * \lfloor \frac{\SYMBOL_{GTS}}{\SYMBOL_A(116)} \rfloor + max\{0, \frac{\SYMBOL_{GTS} \mod \SYMBOL_{A}(116) - (\SYMBOL_H + \SYMBOL_{\ACK} + \SYMBOL_{IFS})}{2}\}, \\
	G_G &= 116 * \lfloor \frac{\SYMBOL_{GTS}}{\SYMBOL_G(116)} \rfloor + max\{0, \frac{\SYMBOL_{GTS} \mod \SYMBOL_G(116) - (\SYMBOL_H +  \SYMBOL_{IFS} - \TURNAROUNDTIME)}{2}\},
\end{align}
with the first part calculating the goodput of packets with a maximum payload of 116 bytes and the second part calculating the maximum payload of a packet that can be sent in the remaining free symbols of a GTS. 

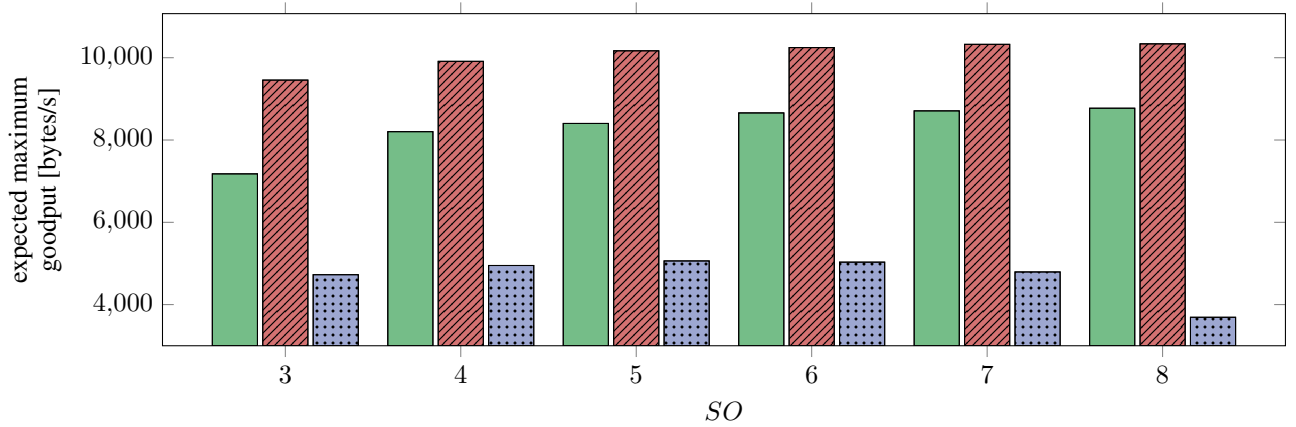
\begin{figure}
    \centering
    \begin{tikzpicture}
\begin{axis}[
    barplot, 
    xlabel = {\SO},
    ylabel = {expected maximum\\goodput [bytes/s]},  
    legend columns = 3,
    ymin = 3000,
    enlarge x limits = 0.14,
]

\foreach \x in {ack, gack, gack_ieee} {
    \addplot+ table[x=sos, y=goodput_\x, col sep=comma] {plots/theoretical/goodput.csv};
}

\end{axis}
\end{tikzpicture} 
    \caption{Expected maximum goodput using regular ACKs and GACKs.}
    \label{fig:theoretical_goodput}
\end{figure}

Fig.~\ref{fig:theoretical_goodput} shows the expected maximum goodput of GACKs ($\mathbb{G}_G$) and regular ACKs ($\mathbb{G}_A$) in bytes for different \SO and $\MO=8$. As one can see, the expected goodput using the proposed GACK schemes is between ~30\% for $\SO=3$ and ~17\% for $\SO=8$ higher than for regular ACKs. For $\SO>8$, the performance difference between both schemes is constant. \GACKIEEE reaches a significant lower goodput than the other schemes because two slots have to be allocated for every transmission. Additionally, the goodput decreases for a decreasing difference between \MO and \SO because two GTS are always allocated for GACK 1 and GACK 2 and hence unusable for regular data transmission.

\section{Evaluation}
To determine whether GACKs are a universally valid alternative to regular ACKs, the schemes are compared in three scenarios. These constitute a worst-case, best-case and average-case scenario for the GACKs which are further described in Sect.~\ref{sec:evaluation_worst}, \ref{sec:evaluation_best} and \ref{sec:evaluation_average}, respectively. Together, these scenarios allow for precise prediction of the GACKs' performance and to identify promising application scenarios. 

The evaluation is conducted using OMNeT++ \cite{varga2008overview}, an event-based simulator, and openDSME \cite{Telematik_Kauer_2019_Diss}, an open-source implementation of IEEE 802.15.4 DSME. The following sections introduce different network topologies with \NODES nodes $\NODE_{0}, \dots, \NODE_{\NODES-1}$ where every node $\NODE_{i\neq 0}$ generates packets with at interval $\tau$ according to a Poisson distribution and sends them to the sink $\NODE_{0}$. The relevant configuration parameters are listed in Tbl.~\ref{tbl:evaluation_paramters}.

\begin{table}
	\centering
	\caption{Parameters for the evaluation of GACKs in different scenarios.}
	\label{tbl:evaluation_paramters}
	\begin{tabular}{c|c|c|c|c|c|c|c}
		& \SO & \MO & \BO & \GAO & topology & nodes & \PAYLOAD \\\hline
		worst-case & 3 & 6 & 8 & 6 & line & 10 &  116 Byte \\ 
		best-case & 4 & 7 & 7 & 7 & star & 19 & 1 Byte \\
		average-case & 4 & 6 & 8 & 6 & tree & 31 &  1 Byte - 116 Byte 
	\end{tabular}	
\end{table} 

\subsection{Worst-case Analysis} \label{sec:evaluation_worst}
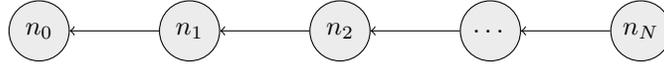
\begin{figure}
    \centering
    \begin{tikzpicture}[node/.style={draw, circle, fill=gray!15, inner sep=0pt, minimum width=.8cm}]
        \node[node] at (0, 0) (n0) {$\NODE_{0}$};
        \node[node] at (2, 0) (n1) {$\NODE_{1}$};
        \node[node] at (4, 0) (n2) {$\NODE_{2}$};
        \node[node] at (6, 0) (n3) {$\dots$};
        \node[node] at (8, 0) (n4) {$\NODE_{N}$};
        \draw[->] (n4) -- (n3);
        \draw[->] (n3) -- (n2);
        \draw[->] (n2) -- (n1);
        \draw[->] (n1) -- (n0);
    \end{tikzpicture}
    \caption{Line topology for the worst-case analysis of GACKs.} 
    \label{fig:evaluation_line_topology}
\end{figure}

GACKs are designed for dense networks where multiple nodes can be acknowledged at once utilizing a single GACK. The opposite of this situation is a line topology as shown in Fig.~\ref{fig:evaluation_line_topology}, where each node can only communicate with its immediate child and parent. Therefore, GACKs only acknowledge a single node at a time, constituting a worst-case scenario. Several packets of a single node can still be acknowledged at once. The following section evaluates GACKs in such a worst-case scenario with $\NODES=10$. Additionally, only a single packet with maximum payload is transmitted per GTS (see Tbl.~\ref{tbl:evaluation_paramters}), impeding the usefulness of GACKs even further.

\subsubsection{Results} \label{sec:evaluation_worst_ideal}
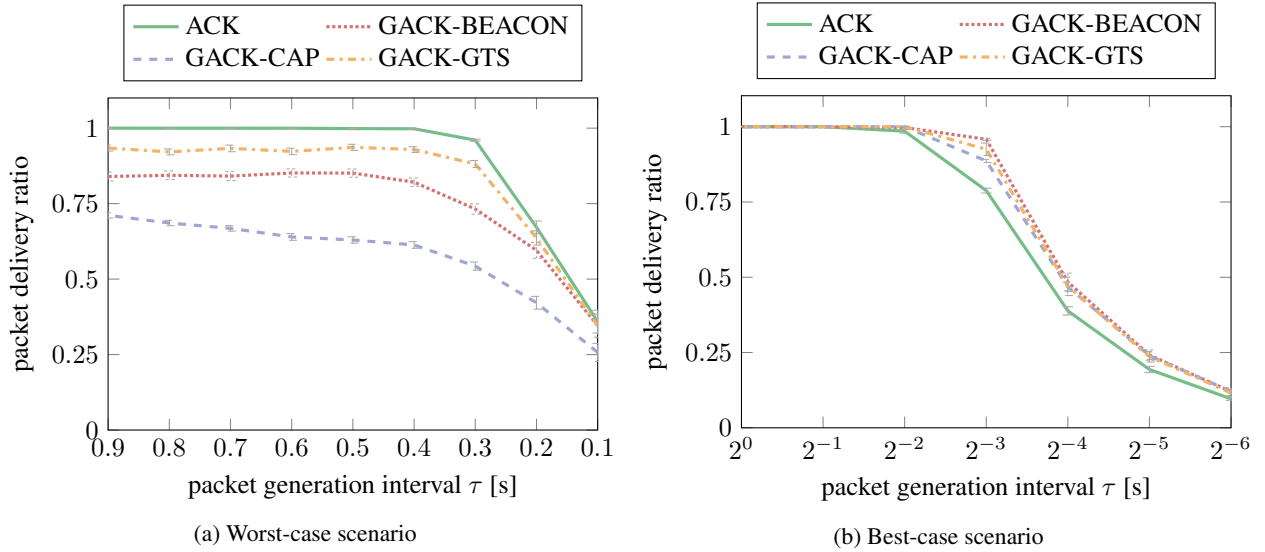
\begin{figure}
    \begin{subfigure}{.49\textwidth}
        \centering
        \begin{tikzpicture}
\begin{axis}[
    lineplot, 
    xlabel = {packet generation interval \PGI [s]},
    ylabel = {packet delivery ratio},  
    legend entries = {ACK, GACK-BEACON, GACK-CAP, GACK-GTS},
    legend style = {at={(0.5,1.05)}, anchor=south},
    legend columns = 2,
    ytick = {0, 0.25, 0.5, 0.75, 1.0},
    x dir=reverse
]

\foreach \x in {ACKPRR:last, BEACONPRR:last, CAPPRR:last, GTSPRR:last} {
    \addplot+ table[x=run, y=\x, col sep=comma, y error=\x_conf] {plots/worst/ideal/data_worst.csv};
}

\end{axis}
\end{tikzpicture} 
        \caption{Worst-case scenario}
        \label{fig:gack_worst_ideal_prr}
    \end{subfigure}
    \hfill
    \begin{subfigure}{.49\textwidth}
        \centering
        \begin{tikzpicture}
\begin{axis}[
    lineplot, 
    xlabel = {packet generation interval \PGI [s]},
    ylabel = {packet delivery ratio},  
    legend entries = {ACK, GACK-BEACON, GACK-CAP, GACK-GTS},
    legend style = {at={(0.5,1.05)}, anchor=south},
    legend columns = 2,
    ytick = {0, 0.25, 0.5, 0.75, 1.0},
    x dir = reverse,    
    xmode=log, 
    log basis x={2}, 
]

\foreach \x in {ACKPRR:last, BEACONPRR:last, CAPPRR:last, GTSPRR:last} {
    \addplot+ table[x=run, y=\x, col sep=comma, y error=\x_conf] {plots/best/ideal/data_best.csv};
}

\end{axis}
\end{tikzpicture} 
        \caption{Best-case scenario}
        \label{fig:gack_best_ideal_prr}
    \end{subfigure}
    \caption{Average PDR for decreasing \PGI using different ACK schemes.}
    \label{fig:evaluation_prr}
\end{figure}

Fig.~\ref{fig:gack_worst_ideal_prr} shows the \textit{packet delivery ratio} (PDR) using different GACK schemes and regular ACKs over a decreasing \PGI. As one can see, regular ACKs perform best while \GACKCAP performs worst. That is because GACKs are transmitted in the CAP using CSMA/CA which suffers from hidden-node problems so that GACKs are frequently lost. For \GACKBEACON, a minimum of $\BO=8$ is required to enable sufficiently many beacon slots for all nodes to join the network. However, with a maximal queue length of 22 packets in openDSME, many packets are dropped because packets are not acknowledged fast enough. At last, \GACKGTS has the overhead of requiring one \GACKGTS per node so that less GTS can be used for regular data packets.   

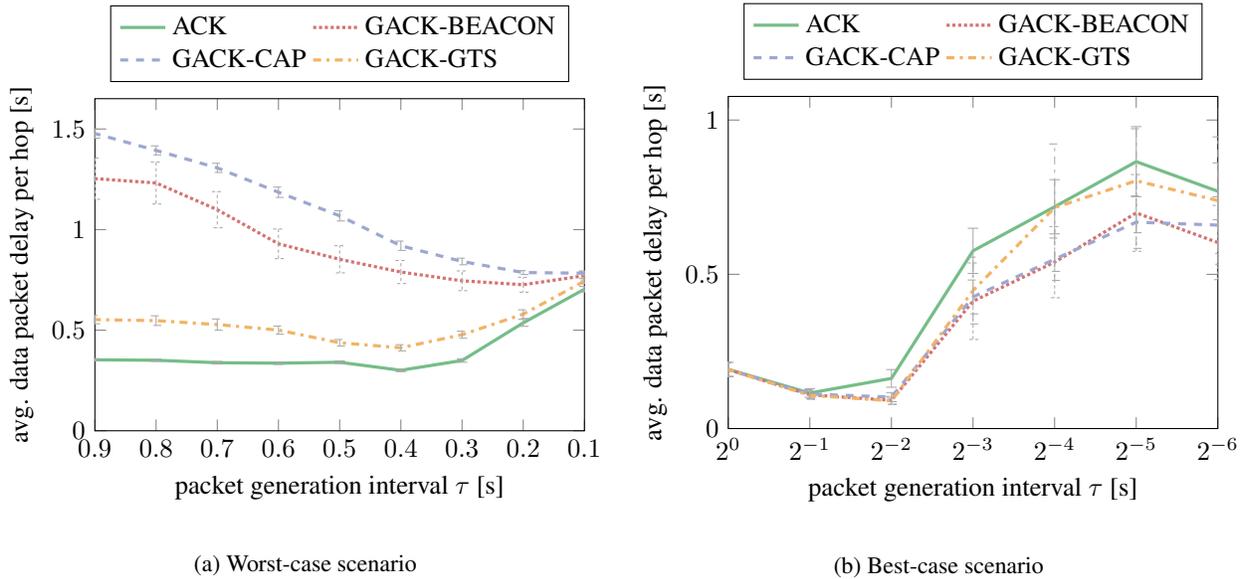
\begin{figure}
    \begin{subfigure}{.49\textwidth}
        \centering
        \begin{tikzpicture}
\begin{axis}[
    lineplot, 
    xlabel = {packet generation interval \PGI [s]},
    ylabel = {avg. data packet delay per hop [s]},  
    legend entries = {ACK, GACK-BEACON, GACK-CAP, GACK-GTS},
    legend style = {at={(0.5,1.05)}, anchor=south},
    legend columns = 2,
    x dir=reverse
]

\def\stat{sinkRcvdPkDelay:mean}
\foreach \x in {ACK, BEACON, CAP, GTS} {
    \addplot+ table[x=run, y=\x\stat, col sep=comma, y error=\x\stat_conf] {plots/worst/ideal/data_worst.csv};
}

\end{axis}
\end{tikzpicture} 
        \caption{Worst-case scenario}
        \label{fig:gack_worst_ideal_data_delay}
    \end{subfigure}
    \hfill
    \begin{subfigure}{.49\textwidth}
        \centering
        \begin{tikzpicture}
\begin{axis}[
    lineplot, 
    xlabel = {packet generation interval \PGI [s]},
    ylabel = {avg. data packet delay per hop [s]},  
    legend entries = {ACK, GACK-BEACON, GACK-CAP, GACK-GTS},
    legend style = {at={(0.5,1.05)}, anchor=south},
    legend columns = 2,
    x dir = reverse, 
    xmode = log, 
    log basis x = {2},
]

\foreach \x in {ACKsinkRcvdPkDelay:mean, BEACONsinkRcvdPkDelay:mean, CAPsinkRcvdPkDelay:mean, GTSsinkRcvdPkDelay:mean} {
    \addplot+ table[x=run, y=\x, col sep=comma, y error=\x_conf] {plots/best/ideal/data_best.csv};
}

\end{axis}
\end{tikzpicture} 
        \caption{Best-case scenario}
        \label{fig:gack_best_ideal_data_delay}
    \end{subfigure}
    \caption{Average end-to-end delay of data packets sent during the CFP for decreasing \PGI. The end-to-end delay of each node is normalized by the number of hops towards the sink.}
    \label{fig:evaluation_data_delay}
\end{figure}

The average data packet delay per hop, i.e., the time between the generation of a data packet at a node and its reception at the sink normalized by the number of hops between the node and the sink, is shown in Fig.~\ref{fig:gack_worst_ideal_data_delay}. As expected, the delay of \GACKCAP and \GACKBEACON is high because GACKs frequently collide using former scheme and the \ACK interval is to high using the latter scheme. \GACKGTS performs slightly worse than regular ACKs due to the \GACKGTS overhead. 

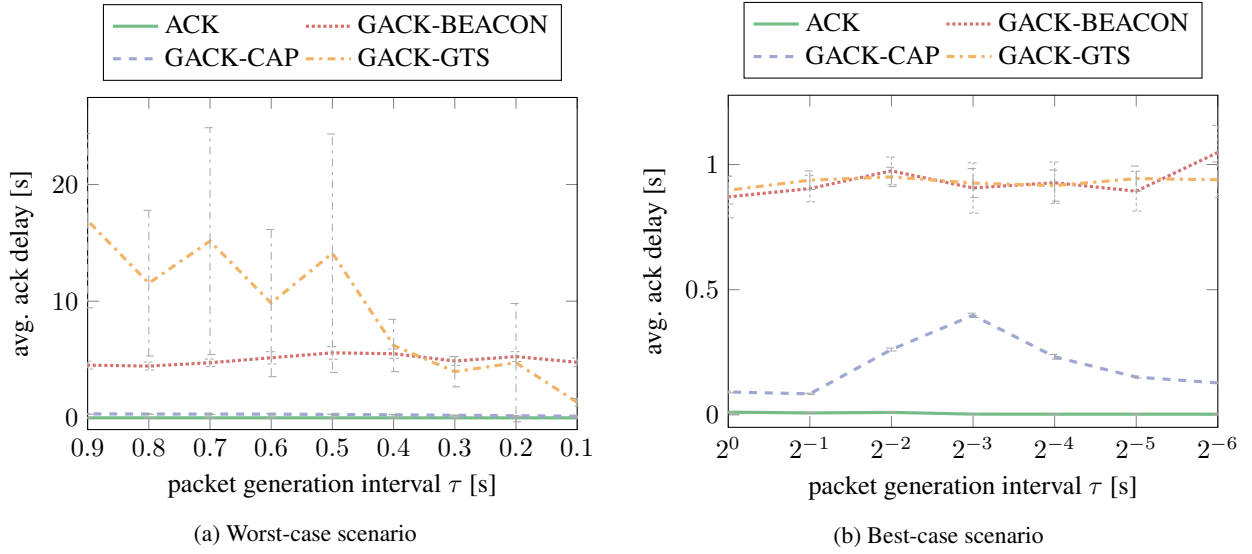
\begin{figure}
    \begin{subfigure}{.49\textwidth}
        \centering
        \begin{tikzpicture}
\begin{axis}[
    lineplot, 
     xlabel = {packet generation interval \PGI [s]},
    ylabel = {avg. ack delay [s]},  
    legend entries = {ACK, GACK-BEACON, GACK-CAP, GACK-GTS},
        legend style = {at={(0.5,1.05)}, anchor=south},
    legend columns = 2,
    x dir=reverse,
    ymin = -1
]

\def\stat{vec_cfpAckDelay:mean}
\foreach \x in {ACK, BEACON, CAP, GTS} {
    \addplot+ table[x=run, y=\x\stat, col sep=comma, y error=\x\stat_conf] {plots/worst/ideal/data_worst.csv};
}

\end{axis}
\end{tikzpicture} 
        \caption{Worst-case scenario}
        \label{fig:gack_worst_ideal_ack_delay}
    \end{subfigure}
    \hfill    
    \begin{subfigure}{.49\textwidth}
        \centering
        \begin{tikzpicture}
\begin{axis}[
    lineplot, 
    xlabel = {packet generation interval \PGI [s]},
    ylabel = {avg. ack delay [s]},  
    legend entries = {ACK, GACK-BEACON, GACK-CAP, GACK-GTS},
        legend style = {at={(0.5,1.05)}, anchor=south},
    legend columns = 2,
    x dir = reverse, 
    xmode = log, 
    log basis x = {2},
    ymin = -0.05
]

\foreach \x in {ACKvec_cfpAckDelay:mean, BEACONvec_cfpAckDelay:mean, CAPvec_cfpAckDelay:mean, GTSvec_cfpAckDelay:mean} {
    \addplot+ table[x=run, y=\x, col sep=comma, y error=\x_conf] {plots/best/ideal/data_best.csv};
}

\end{axis}
\end{tikzpicture} 
        \caption{Best-case scenario}
        \label{fig:gack_best_ideal_ack_delay}
    \end{subfigure}
    \caption{Average time between sending a data packet and receiving the according acknowledgment for decreasing \PGI.}
    \label{fig:evaluation_ack_delay}
\end{figure}

Fig.~\ref{fig:gack_worst_ideal_ack_delay} shows the average ACK delay, i.e., the time from transmitting a data packet to receiving the according acknowledgment. \GACKBEACON and \GACKGTS provide the worst performance due to their large acknowledgment interval, however, the delay of \GACKGTS is unstable for \PGI from 0.9s to 0.4s. That is because a \GACKGTS is allocated through the regular GTS-handshake. Due to hidden-node problems many of these handshake-messages fail and thus the allocation of \GACKGTS is delayed.   

\begin{figure}
    \begin{subfigure}[b]{.49\textwidth}
        \centering
        \begin{tikzpicture}
\begin{axis}[
    lineplot, 
    xlabel = {interference interval [s]},
    ylabel = {packet delivery ratio},  
    legend entries = {ACK, GACK-BEACON, GACK-CAP, GACK-GTS},
    ytick = {0, 0.25, 0.5, 0.75, 1.0},
    legend style = {at={(0.5,1.05)}, anchor=south},
    legend columns = 2,
    x dir=reverse
]

\foreach \x in {ACKPRR:last, BEACONPRR:last, CAPPRR:last, GTSPRR:last} {
    \addplot+ table[x=run, y=\x, col sep=comma, y error=\x_conf] {plots/worst/jammed/data_worst_jammed.csv};
}

\end{axis}
\end{tikzpicture} 
        \caption{Worst-case scenario}
        \label{fig:gack_worst_jammed_prr}
    \end{subfigure}
    \hfill
    \begin{subfigure}[b]{.49\textwidth}
        \centering
        \begin{tikzpicture}
\begin{axis}[
    lineplot,
    xlabel = {interference interval [s]},
    ylabel = {packet delivery ratio},  
    legend entries = {ACK, GACK-BEACON, GACK-CAP, GACK-GTS},
    legend style = {at={(0.5,1.05)}, anchor=south},
    legend columns = 2,
    ytick = {0, 0.25, 0.5, 0.75, 1.0},
    x dir = reverse, 
]

\foreach \x in {ACKPRR:last, BEACONPRR:last, CAPPRR:last, GTSPRR:last} {
    \addplot+ table[x=run, y=\x, col sep=comma, y error=\x_conf] {plots/best/jammed/data_best_jammed.csv};
}

\end{axis}
\end{tikzpicture} 
        \caption{Best-case scenario}
        \label{fig:gack_best_jammed_prr}
    \end{subfigure}
    \caption{PDR of different ACK schemes under external interference of a packet with maximum \PAYLOAD and decreasing interval.}
    \label{fig:evaluation_jammed_prr}
\end{figure}
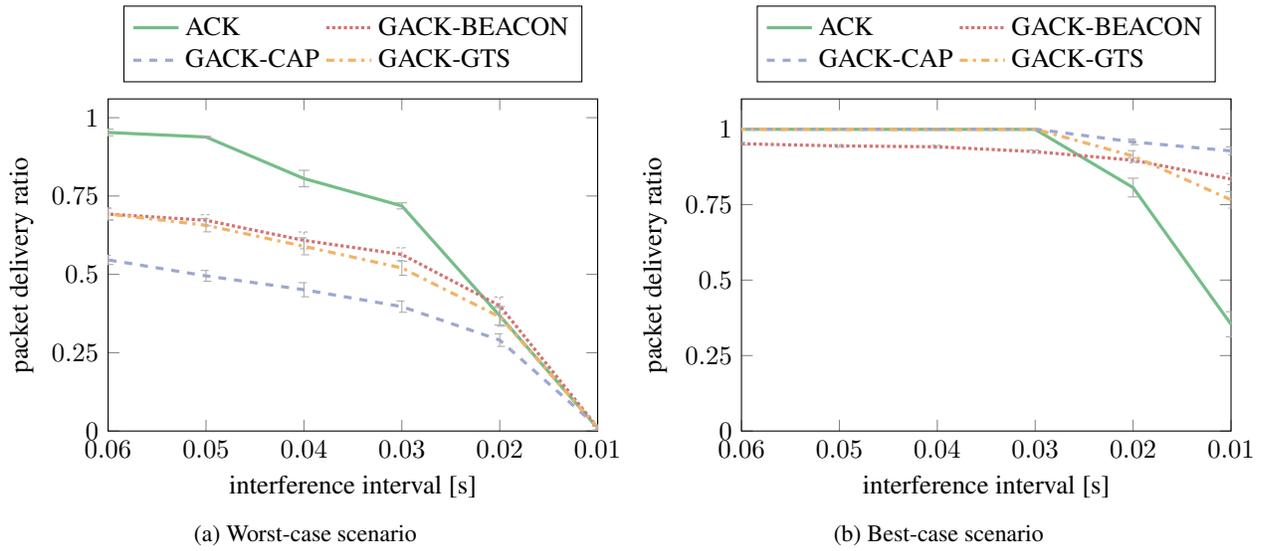

At last, Fig.~\ref{fig:gack_worst_jammed_prr} depicts the PDR of the different ACK schemes in a more realistic scenario, i.e., under external interference of a packet with maximum payload and varying interference interval. Interference lasts for about 5ms. For sake of simplicity, the interference occurs on all channels simultaneously and \PGI is fixed to 1.0 s. The relation between the ACK schemes is similar as before with one exception, \GACKBEACON performs slightly better than the other schemes, including regular ACKs, for an interference interval of 20 ms and lower.
 
\subsection{Best-case Analysis} \label{sec:evaluation_best}
\begin{figure}
    \centering
    \includegraphics{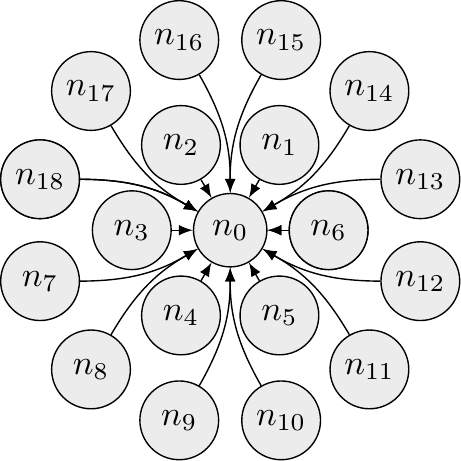}
    \caption{Star topology for the best-case analysis of GACKs.}
    \label{fig:evaluation_star_topology}
\end{figure}

As described in Sect.~\ref{sec:evaluation_worst}, GACKs are designed for dense networks with several nodes within communication range of the sink. Thus, the star topology depicted in Fig.~\ref{fig:evaluation_star_topology} constitutes a best-case scenario for GACKs as multiple nodes can be acknowledged at once. The evaluated network consists of 19 nodes. Sect.~\ref{sec:evaluation_best_ideal} analyses the performance of GACKs in this best-case scenario with and without external interference and tries to find an upper performance bound in terms of throughput, data packet delay and acknowledgment delay. A small payload size of 1 Byte is chosen while \MO and \SO are chosen in a way that many GTS can be acknowledged using a single GACK. 

\subsubsection{Results} \label{sec:evaluation_best_ideal}
Fig.~\ref{fig:gack_best_ideal_prr} shows the PDR in a best-case scenario, where the PDR of regular ACKs decreases slightly faster for decreasing \PGI than the GACK schemes. All GACK schemes perform almost the same, where GACK beacon performs best because the beacon interval is low and there is no additional overhead during the CFP. 

The same applies to the average data packet delay, depicted in Fig.~\ref{fig:gack_best_ideal_data_delay}, where regular ACKs result in the largest delay because less packets can be transmitted per GTS and hence packets remain in the queue longer. \GACKCAP and \GACKBEACON achieve the lowest delay, where \GACKGTS introduces a slightly higher delay due to the allocation through the 3-way handshake. At last, the  acknowledgment delay in Fig.~\ref{fig:gack_best_ideal_ack_delay} behaves similar to the worst-case scenario. 

Under the influence of external interference, more packets can be transmitted using GACKs for small \PGI, as presented in Fig.~\ref{fig:gack_best_jammed_prr}. That is because more packets can be transmitted per GTS so that a single lost packet has less influence. 

In general, the performance difference between regular ACKs and GACKs is smaller than in the worst-case scenario, discussed in Sect.~\ref{sec:evaluation_worst}, indicating that GACKs can provide improvements in optimal scenarios but also impose great risks as their worst-case performance is much worse. 

\subsection{Average-case Analysis} \label{sec:evaluation_average}
At last, GACKs are evaluated in an average-case scenario, which we exemplary choose as a binary-tree topology with 31 nodes. Here, packets are generated with a random payload between 1 Byte and 116 Bytes to reflect common use cases. It should be noted that a binary-tree contains many hidden-node problems similar to the line topology but also offers the possibility to acknowledge two nodes using a single GACK. 

\subsubsection{Results}
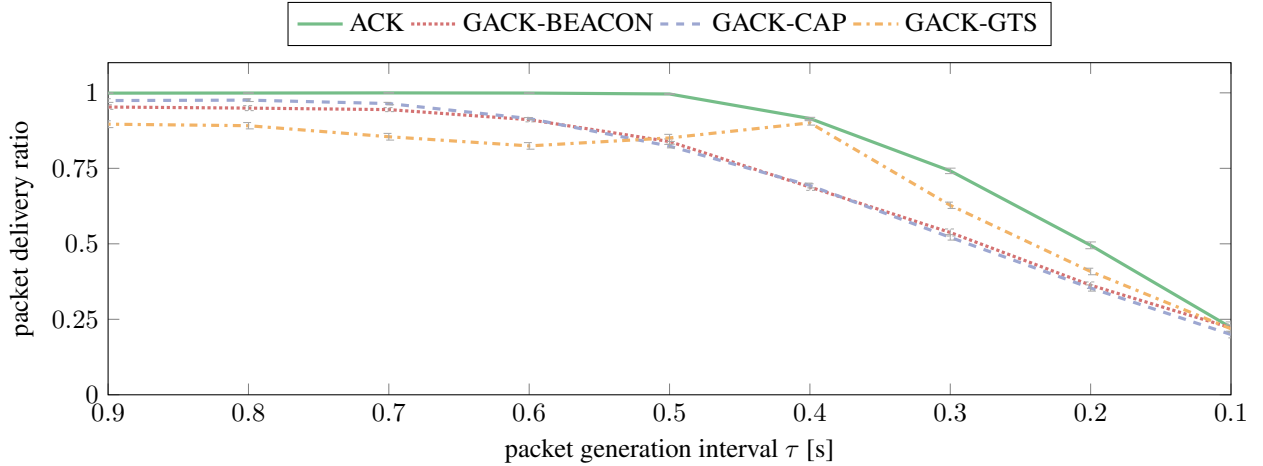
\begin{figure}
	\centering
	\begin{tikzpicture}
\begin{axis}[
    lineplot, 
    xlabel = {packet generation interval \PGI [s]},
    ylabel = {packet delivery ratio},  
    legend entries = {ACK, GACK-BEACON, GACK-CAP, GACK-GTS},
    legend style = {at={(0.5,1.05)}, anchor=south},
    legend columns = 4,
    ytick = {0, 0.25, 0.5, 0.75, 1.0},
    x dir = reverse,    
]

\foreach \x in {ACKPRR:last, BEACONPRR:last, CAPPRR:last, GTSPRR:last} {
    \addplot+ table[x=run, y=\x, col sep=comma, y error=\x_conf] {plots/avg/ideal/data_avg.csv};
}

\end{axis}
\end{tikzpicture} 
	\caption{Packet delivery ratio for different ACK schemes for decreasing \PGI in a binary-tree topology.}
	\label{fig:evaluation_avg_prr}
\end{figure}

Fig.~\ref{fig:evaluation_avg_prr} shows the PDR of the different GACK schemes and regular ACKs for decreasing \PGI. Here, regular ACKs achieve the highest PDR while the performance of the GACK schemes highly depends on \PGI. \GACKGTS suffers from the initial overhead of one \GACKGTS per node and hence only achieves about 90\% PDR for \PGI between 0.9 s and 0.4 s. However, in comparison to \GACKCAP and \GACKBEACON which achieve a PDR of about 97\% and 98\% initially, respectively, \GACKGTS manages to maintain the same PDR for higher \PGI, where the performance of \GACKCAP and \GACKBEACON falls off because of increased management traffic in the CAP and a too large acknowledgment interval using  beacons. All GACK schemes perform worse than regular ACKs. 

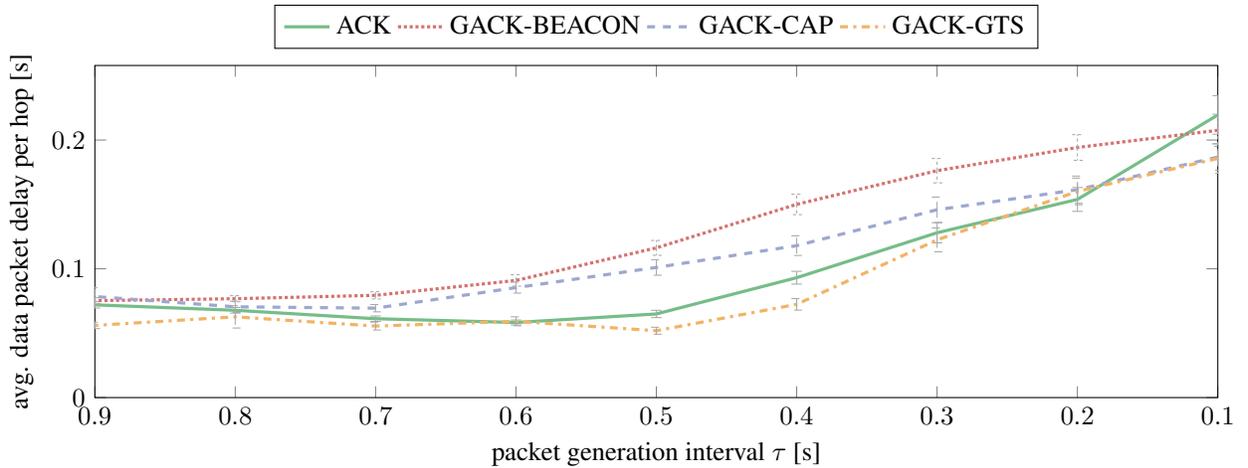
\begin{figure}
	\centering
	\begin{tikzpicture}
\begin{axis}[
    lineplot, 
    xlabel = {packet generation interval \PGI [s]},
    ylabel = {avg. data packet delay per hop [s]},  
    legend entries = {ACK, GACK-BEACON, GACK-CAP, GACK-GTS},
    legend style = {at={(0.5,1.05)}, anchor=south},
    legend columns = 4,
    x dir = reverse, 
]

\foreach \x in {ACKsinkRcvdPkDelay:mean, BEACONsinkRcvdPkDelay:mean, CAPsinkRcvdPkDelay:mean, GTSsinkRcvdPkDelay:mean} {
    \addplot+ table[x=run, y=\x, col sep=comma, y error=\x_conf] {plots/avg/ideal/data_avg.csv};
}

\end{axis}
\end{tikzpicture} 
	\caption{Average end-to-end delay of data packets sent during the CFP for decreasing \PGI. The end-to-end delay of each node is normalized by the number of hops towards the sink.}
	\label{fig:evaluation_avg_data_delay}
\end{figure}

On the other hand, \GACKGTS results in the lowest data delay and even outperforms regular ACKs, as shown in Fig.~\ref{fig:evaluation_avg_data_delay}, as more packets can be transmitted per GTS. \GACKBEACON and \GACKCAP exhibit the highest delay for lower \PGI because less packets are transmitted successfully and the queue starts to fill up. 

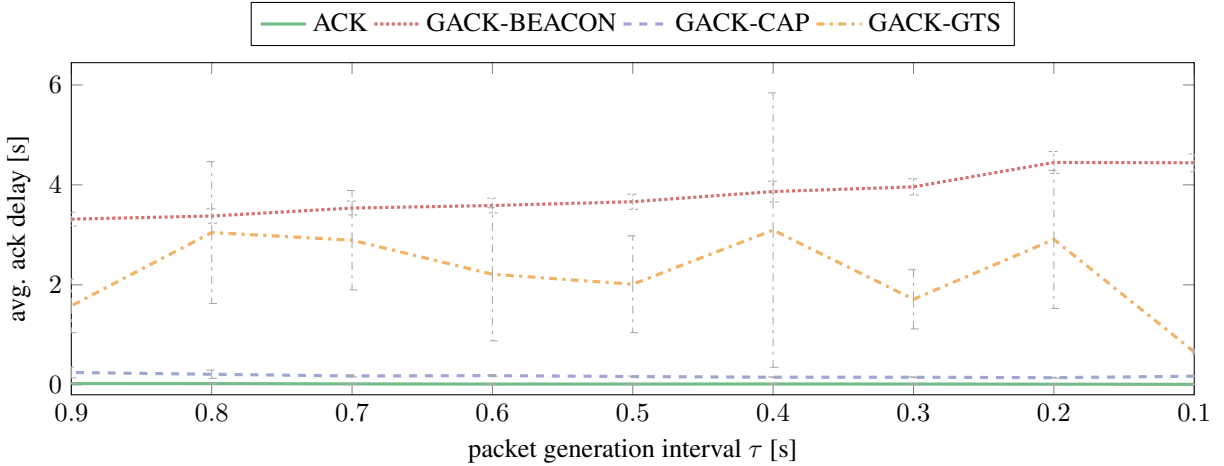
\begin{figure}
	\centering
	\begin{tikzpicture}
\begin{axis}[
    lineplot, 
    xlabel = {packet generation interval \PGI [s]},
    ylabel = {avg. ack delay [s]},  
    legend entries = {ACK, GACK-BEACON, GACK-CAP, GACK-GTS},
    legend style = {at={(0.5,1.05)}, anchor=south},
    legend columns = 4,
    x dir = reverse, 
    ymin = -0.2
]

\foreach \x in {ACKvec_cfpAckDelay:mean, BEACONvec_cfpAckDelay:mean, CAPvec_cfpAckDelay:mean, GTSvec_cfpAckDelay:mean} {
    \addplot+ table[x=run, y=\x, col sep=comma, y error=\x_conf] {plots/avg/ideal/data_avg.csv};
}

\end{axis}
\end{tikzpicture} 
	\caption{Average time between sending a data packet and receiving the according acknowledgment for decreasing \PGI.}
	\label{fig:evaluation_avg_ack_delay}
\end{figure}

The acknowledgment delay, depicted in Fig.~\ref{fig:evaluation_avg_ack_delay},  behaves similar to the best-case and worst-case scenarios, where regular ACKs achieve a minimal acknowledgment delay while \GACKCAP acknowledges packets after every superframe and thus also achieves a relatively low delay. For the given network a high \BO is required so that the acknowledgment delay of \GACKBEACON is high. However, one should be aware that the acknowledgment delay is only relevant if packets are lost and retransmitted. Also, the acknowlegement delay contributes to the data packet delay where retransmitted packets show a high delay which includes the acknowledgment delay. 

\subsubsection{FIT IoT-LAB}
\begin{figure}
    \centering
    \includegraphics{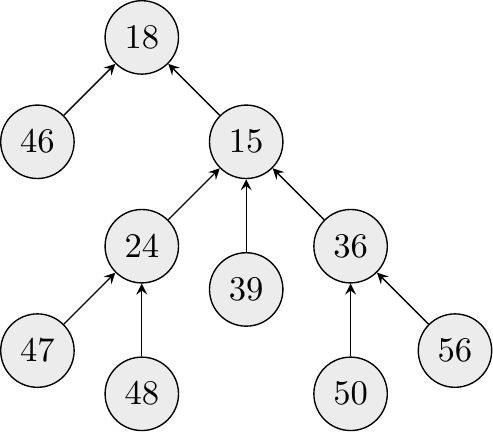}
    \caption{Tree topology utilized for experiments in the FIT IoT-LAB. The numbers indicate the node IDs in the strasbourg testbed.}
    \label{fig:iotlab_topology} 
\end{figure}

To verify the applicability of GACKs in a scenario with physical nodes, we conducted energy-measurements in the \textit{strasbourg} testbed of the FIT IoT-LAB \cite{adjih_iotlab}. Experiments are conducted in a tree topology with 10 nodes and depth three, as shown in Fig.~\ref{fig:iotlab_topology}, generated by the algorithm proposed in \cite{Telematik_Kauer_2019_Diss}. The target hardware is an M3 Open Node with a 72 MHz, 32-bit Cortex M3 CPU, 64 KB RAM, and 256 KB ROM. According to \cite{Telematik_Kauer_2019_Diss}, transmission power was configured to -3 dBm and transceiver sensitivity to -60 dBm. Experiments are repeated 10 times, where 1000 packets with $\tau = 0.5 s$ are evaluated.   

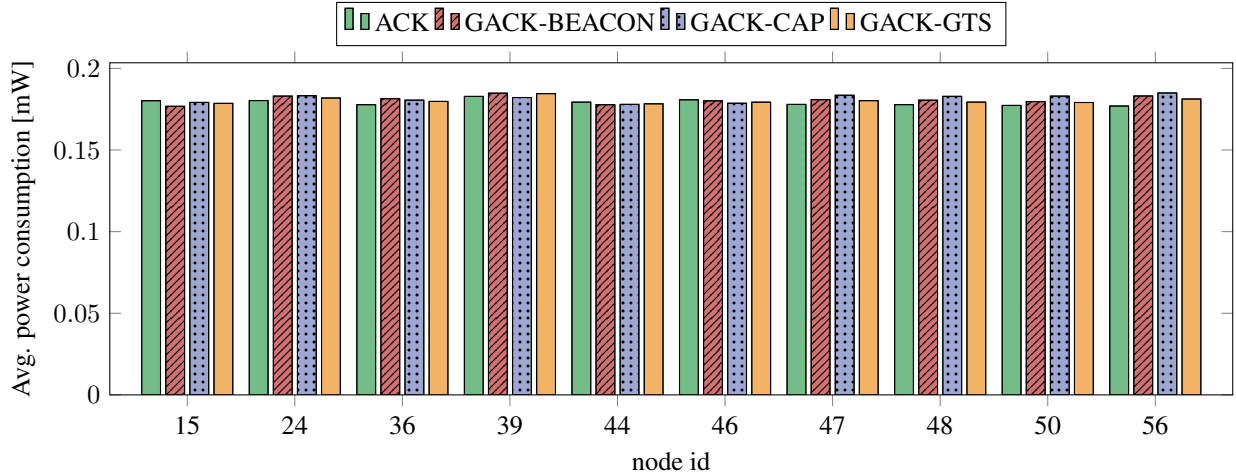
\begin{figure}
    \centering
    \begin{tikzpicture}
\begin{axis}[
    barplot, 
    xlabel = {node id},
    ylabel = {Avg. power consumption [mW]},  
    enlarge x limits = 0.08,
    symbolic x coords = {15, 24, 36, 39, 44, 46, 47, 48, 50, 56},
    bar width=0.25cm,
    legend entries = {ACK, GACK-BEACON, GACK-CAP, GACK-GTS},
    legend style = {at={(0.5,1.05)}, anchor=south},
    legend columns = 4,
]

\foreach \x in {ACK, GACK-BEACON, GACK-CAP, GACK-GTS} {
    \addplot+ table[x=id, y=\x power, col sep=comma] {plots/avg/iotlab/data_power.csv};
}

\end{axis}
\end{tikzpicture} 
    \caption{Average power consumption for different nodes in the FIT IoT-LAB.}
    \label{fig:iotlab_power}
\end{figure}

Fig.~\ref{fig:iotlab_power} shows the average power consumption of all nodes in the tree topology shown in Fig.~\ref{fig:iotlab_topology}. Opposing our initial hypothesis, there is no significant difference in the power consumption between GACKs and regular ACKs. That is due to two reasons: In high-traffic scenarios, GACKs are likely to transmit additional data packets per GTS which offer no benefit over transmitting ACKs as the switching power consumption is close to the transmission power consumption. In low-traffic scenarios, on the other hand, power consumption is dominated by the CAP which makes up 56.25\% of a superframe and in which the receiver has to be turned on to exchange management messages. Thus, saving only a few ACKs during a GTS does not constitute any power saving. 

\subsection{Discussion}
Summarizing our previous results, one has to say that GACKs do not provide significant benefits in terms of throughput, data packet delay and energy over the tested scenarios to justify their usage over regular acknowledgments. On the one hand, they can provide small performance increases in term of throughput and reliability in optimal environments, e.g., in star topologies where many nodes can be acknowledged at once, but on the other hand they suffer from significant performance loss in comparison to regular ACKs in anything but optimal scenarios. Additionally, regular ACKs prove to be superior in environments with strong interference because many packets have to be buffered until retransmission using GACKs resulting in frequent packet drops. Thus, we could not identify GACKs as a universally valid alternative to regular ACKs and, even though star topologies are quite common in industrial setups, the minor performance gain in these topologies does not justify the usage of GACKs. Even by transmitting GACKs in multiple ways, no solution was able to verify results from literature \cite{sahoo2017novel, alderisi2015simulative} where GACKs are handled as a valid alternative to regular ACKs.   

Above statement is reinforced by the high implementation effort for GACKs which require additional message definitions, alteration of the transmission mechanism, and adapted GTS and transceiver management, i.e., large deviations from the official standard. Furthermore, GACKs impose a large memory overhead for storing data packets until they are acknowledged. Consequently, we conclude that the achieved performance benefits of GACK are usually not worth the additional implementation effort. At last, it should be said that GACKs work best in single hop networks, e.g. star topologies, with many nodes. DSME is explicitly built for scalable multi-hop networks and such scenarios keep DSME from performing best by definition limiting the effectiveness of GACKs in DSME even further. If a GACK scheme should be used, \GACKCAP offers the best trade-off between throughput and acknowledgment delay but heavily suffers from hidden node problems. In scenarios with many hidden nodes problems, \GACKGTS offers a valid alternative.

\section{Conclusion}
In this work, we evaluate whether group acknowledgments (GACKs), which aggregate acknowledgments for data packets and broadcast them to all nodes in the neighborhood using a single message, offer a valid alternative to the direct acknowledgment of data packets. For this, three novel ways of transmitting GACKs are evaluated and compared in a worst-case line topology, best-case star topology and average-case tree topology. Results indicate that GACKs can provide a 20\% higher PDR than regular ACKs in a best-case scenario but their performance is significantly worse than regular ACKs in all other scenarios. Thus, we believe that GACKs do not constitute a valid alternative to the direct acknowledgment of data packets, especially because they also impose a large storage overhead for packets until they are acknowledgement. From all tested GACK schemes, transmitting GACKs in the CAP provides the best trade-off between acknowledgment delay and throughput.

\bibliography{document}{}
\bibliographystyle{plain}

\end{document}